\documentclass[twocolumn]{aastex63}
\usepackage{xcolor}

\newcommand{\quasar}{P172+18}

\shorttitle{Resolving The Radio-Loud Quasar P172+18 at $z = 6.82$}
\shortauthors{Momjian et al.}

\begin{document}

\title{RESOLVING THE RADIO EMISSION FROM THE QUASAR P172+18 AT $\lowercase{z} = 6.82$}

\correspondingauthor{Emmanuel Momjian}
\email{emomjian@nrao.edu}

\author[0000-0003-3168-5922]{Emmanuel Momjian}
\affiliation{National Radio Astronomy Observatory, P.O. Box O, Socorro, NM 87801, USA}

\author[0000-0002-2931-7824]{Eduardo Ba\~nados}
\affiliation{{Max Planck Institut f\"ur Astronomie, K\"onigstuhl 17, D-69117, Heidelberg, Germany}}

\author[0000-0001-6647-3861]{Christopher L. Carilli}
\affiliation{National Radio Astronomy Observatory, P.O. Box O, Socorro, NM 87801, USA}

\author[0000-0003-4793-7880]{Fabian Walter}
\affiliation{{Max Planck Institut f\"ur Astronomie, K\"onigstuhl 17, D-69117, Heidelberg, Germany}}

\author[0000-0002-5941-5214]{Chiara Mazzucchelli}\thanks{ESO Fellow}
\affiliation{European Southern Observatory, Alonso de Cordova 3107, Vitacura, Region Metropolitana, Chile}

\accepted{February 11, 2021}

\begin{abstract}
We present high angular resolution imaging of the quasar PSO~J172.3556+18.7734 at $z=6.82$ with the Very Long Baseline Array (VLBA). This source currently holds the record of being the highest redshift radio-loud quasar. These observations reveal a dominant radio source with a flux density of $398.4 \pm 61.4\,\mu$Jy at 1.53\,GHz, a deconvolved size of $9.9 \times 3.5$ mas ($52.5 \times 18.6$\,pc), and an intrinsic brightness temperature of ($4.7 \pm 0.7) \times 10^7$\,K. A weak unresolved radio extension from the main source is also detected at $\sim~3.1\sigma$ level. The total flux density recovered with the VLBA at 1.53\,GHz is consistent with that measured with the Very Large Array (VLA) at a similar frequency. The quasar is not detected at 4.67\,GHz with the VLBA, suggesting a steep spectral index with a limit of $\alpha^{1.53}_{4.67} < -$1.55. The quasar is also not detected with the VLBA at 7.67\,GHz. The overall characteristics of the quasar suggest that it is a very young radio source similar to lower redshift Gigahertz Peaked Spectrum radio sources, with an estimated kinematic age of $\sim~10^3$\,years. The VLA observations of this quasar revealed a second radio source in the field 23\farcs1 away. This radio source, which does not have an optical or IR counterpart, is not detected with the VLBA at any of the observed frequencies. Its non-detection at the lowest observed VLBA frequency suggests that it is resolved out, implying a size larger than $\sim~0\farcs17$. It is thus likely situated at lower redshift than the quasar.
\end{abstract}

\keywords{cosmology: observations --- cosmology: early universe  --- galaxies: high-redshift ---
quasars: individual (PSO~J172.3556+18.7734) --- radio continuum: galaxies --- technique: interferometric}

\section{Introduction} \label{sec:intro}

Quasars with billion-solar-mass black holes have been detected well within the epoch of reionization,  when the universe was less than one Gyr old \citep{maz17,ban18,wang18,yang20}. These objects challenge our understanding of the formation and growth of supermassive black holes (SMBHs). A potential mechanism to grow these SMBHs is through jet-enhanced accretion, which can enable Super-Eddington accretion rates \citep{jk08,vol15}. Therefore, the study of jets on the first quasars could provide key clues about this outstanding question in astrophysics. 

Quasars with strong radio emission are of particular interest given that Very Long Baseline Interferometry (VLBI) radio observations are the only way to investigate their jets on pc-scales. 

In this paper, we present VLBI observations of the recently discovered radio-loud quasar PSO~J172.3556+18.7734 (hereafter \quasar) at $z = 6.823$ \citep{ban21}.
This is currently the only radio source known at $z>6.45$. 
\quasar\ is among the fastest accreting quasars at any redshift with an Eddington ratio of $\sim2.2$, and hosts a supermassive black hole with a mass of $M_{\mathrm{BH}} \sim 2.9 \times 10^{8} M_{\odot}$. Karl G. Jansky Very Large Array (VLA)
observations show that the optical quasar is associated with an unresolved radio source with a 
flux density of $510 \pm 15$\,$\mu$Jy at 1.52\,GHz and $222 \pm 9$\,$\mu$Jy at 2.87\,GHz, and a
size smaller than $1\farcs9 \times 0\farcs87$, or $10.1 \times 4.6$\,kpc \citep{ban21}.
The implied rest frame 1.4\,GHz luminosity density is
$L_{\rm \nu,1.4\,GHz} = (5.8 \pm 0.2) \times 10^{26}$~W~Hz$^{-1}$.

In addition to the quasar, a second radio source was identified with the VLA at an angular distance of 23\farcs1 from the quasar. 
Although the redshift of this ``companion'' source is still unconstrained, the chances of having these two radio-sources at the same distance of the quasar itself is less than $2\%$ based on deep field number counts \citep{ban21}.  
This ``companion'' source has no optical, near-infrared, or mid-infrared counterpart in the currently available shallow data. It is unresolved at 1.52\,GHz with a flux density of $732 \pm 15$\,$\mu$Jy and a size smaller than $1\farcs6 \times 0\farcs69$, and resolved at 2.87\,GHz with a flux density of $432 \pm 20$\,$\mu$Jy and a deconvolved size of $1\farcs3 \times 0\farcs8$ \citep{ban21}. The coordinates of the quasar and the ``companion'' source are listed in Table\,1 of \citet{ban21}.

In Section \ref{obs} we present the 1.53, 4.67, and 7.67\,GHz VLBI  observations and their data reduction. 
In Section \ref{resultsandanalysis} we present the results and analysis for \quasar\ and the ``companion'' source, separately. 
Finally, in Section \ref{disc} we discuss the VLBI results and compare the quasar \quasar\ with other known radio-loud quasars near $z \sim 6$. 
We adopt a flat cosmology with $H_0 = 70
\,\mbox{km\,s}^{-1}$\,Mpc$^{-1}$, $\Omega_M = 0.3$, and
$\Omega_\Lambda = 0.7$.  At the redshift of this quasar, 1\,mas
corresponds to 5.3\,pc.

\section{Observations and Data Reduction} \label{obs}

The VLBI observations of the \quasar\ were carried out at 1.53\,GHz (L-band) on
2019 July 20, and simultaneously at 4.67 and 7.67\,GHz (C-band) on 2019 October 29,
using the Very Long Baseline Array (VLBA) of the
NRAO. At L-band, eight 32\,MHz data channel pairs were
recorded at each station using the ROACH Digital Backend and the
polyphase filterbank (PFB) digital signal-processing algorithm, both with right- and left-hand circular polarizations, and sampled at two bits. The total bandwidth was 256\,MHz centered at 1.53\,GHz. The total observing time at L-band was 6\,hr, with 4.2\,hr on target.

At C-band, which nominally covers the frequency range
3.9--7.9\,GHz, four 64\,MHz data channel pairs were
recorded at each station using the ROACH Digital Backend and the digital downconverter (DDC) signal-processing algorithm, also both with right- and left-hand circular polarizations, and sampled at two bits. Two data channel pairs were tuned near the lower end of the C-band receiver's frequency span, and the other two near its higher end, allowing for simultaneous observations at widely separated frequencies. The total bandwidth per used C-band frequency tuning was 128\,MHz, and the two tunings were centered at 4.67\,GHz and 7.67\,GHz. The total observing time at C-band was also 6\,hr (4.2\,hr on-target time).

At both L- and C-bands, the VLBA observations utilized nodding-style phase referencing using the calibrator J1122+1805, which is separated by $1\rlap{.}{^\circ}78$ from the target
source. The phase referencing cycle time was 3.45\,min:
2.75\,min on the target and 0.7\,min on the calibrator.
The uncertainty in the calibrator's position is
0.08\,mas in right ascension and 0.11\,mas in declination \citep{cha20}. The accuracy of the phase calibrator position is important in phase-referencing observations \citep {WAL99}, because it determines the accuracy of the absolute position of the target. As employed in these observations, phase referencing would preserve absolute astrometric positions to better than $\pm 0\rlap{.}^{''}01$ \citep{FOM99}. The observations also included the calibrator source 4C\,39.25 which was used as a fringe finder and bandpass calibrator. Amplitude calibration was performed using measurements of the antenna gain and the system temperature of each station.

The data were correlated with the VLBA DiFX correlator \citep{DEL11} in Socorro, NM, with 1\,s correlator integration time. Two correlation passes were performed on the data of each
observing session: the first pass was at the position of the quasar \quasar, and the second was at the position of the VLA detected ``companion'' radio source located at an angular distance of 23\farcs1 from the quasar.

Data reduction and analysis were performed using the Astronomical Image Processing System (AIPS: \citealt{G2003}) following standard VLBI data reduction procedures. The phase reference source was self-calibrated and the solutions were applied on the target fields (the quasar and the ``companion''). Deconvolution and imaging were performed using a grid weighting near the mid point between pure natural and pure uniform
(Robust=1 in AIPS task IMAGR).

\begin{figure}
\epsscale{1.2}
\plotone{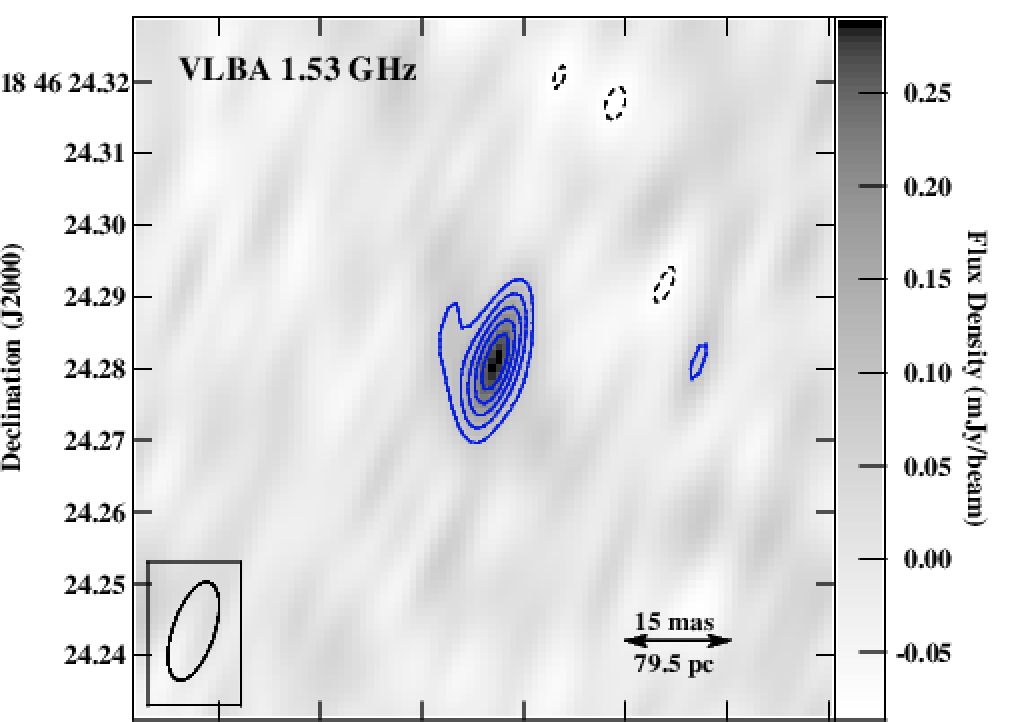}
\caption
{VLBA continuum image of the $z=6.82$ quasar \quasar\ at 1.53\,GHz and $14.3 \times
5.8$~mas resolution (P.~A.=$-18^{\circ}$). The contour levels are at $-3$, 3, 4.5, 6,
7.5, and 9 times the rms noise level, which is
27.5~$\mu$Jy~beam$^{-1}$. The gray-scale range is indicated by the step wedge at the
right side of the image. 
\label{fig:vlba1}}
\end{figure}

\section{Results and Analysis} \label{resultsandanalysis}

\subsection{The Quasar \quasar}
\label{raqso}

\begin{figure*}[ht]
\epsscale{0.9}
\plottwo{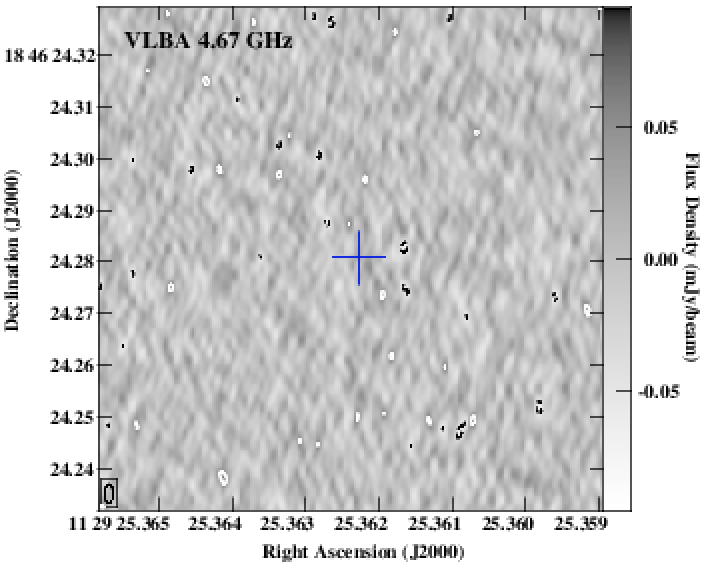}{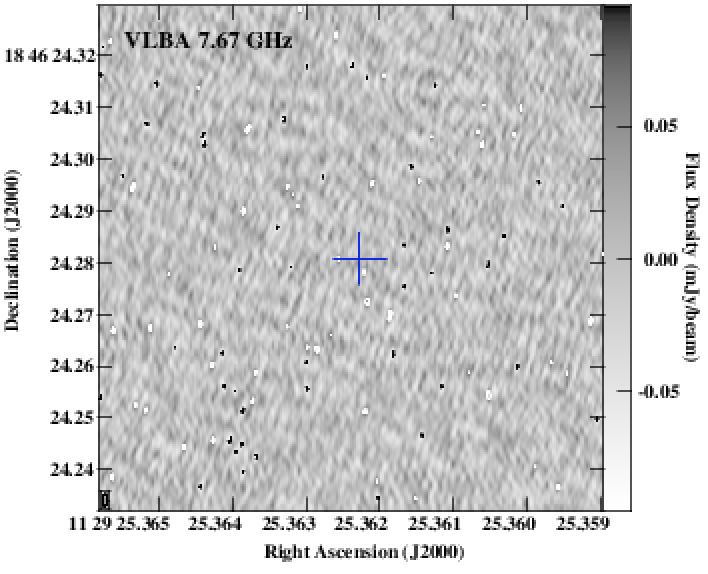}
\caption
{VLBA continuum images of the $z=6.82$ quasar \quasar\ at 4.67 ({\it Left}) and 7.67\,GHz
({\it Right}). Their synthesized beam sizes are $3.64 \times 1.62$~mas (P.~A.=$-2^{\circ}$),
and $2.39 \times 0.98$~mas (P.~A.=$-7^{\circ}$), respectively. 
The contour levels are at $-3$ and 3 times the rms noise level in each image, which is
17.2~$\mu$Jy~beam$^{-1}$ at 4.67\,GHz and 20.4~$\mu$Jy~beam$^{-1}$ at 7.67\,GHz. The
gray-scale range is indicated by the step wedge at the right side of each image. 
The plus sign in each image denotes the VLBA position of the dominant component of \quasar\ at 1.53\,GHz as seen in Figure\,1, which is 
R.A.\,(J2000)=\,11$^{\rm h}$29$^{\rm m}$25.36227$^{\rm s}$, 
Decl.\,(J2000)=\,$+$18$^\circ$46$^\prime$24\farcs2808.
\label{fig:vlba2}}
\end{figure*}

Figure~1 shows the VLBA 1.53\,GHz image of  
\quasar\ at an angular resolution of $14.3 \times
5.8$~mas ($75.8 \times 30.7$\,pc at $z=6.82$) with a position angle of
P.~A.=$-18^{\circ}$. The rms noise in the image is 27.5~$\mu$Jy~beam$^{-1}$.
The observing frequency of 1.53\,GHz corresponds
to a rest frame frequency of 11.96\,GHz.
The image shows a dominant continuum source with a peak flux density of
$289.5 \pm 27.5$~$\mu$Jy~beam$^{-1}$, and a weak radio extension from the main source to the north-east.
Performing a two component 2-dimensional Gaussian  
fit resulted in a resolved component for the dominant radio source with a total flux density of 
$398.4 \pm 61.4$\,$\mu$Jy, and a deconvolved size of 
$9.9 \times 3.5$~mas ($52.5 \times 18.6$\,pc at $z=6.82$). The  
corresponding intrinsic brightness temperature value, i.e., at the rest frame frequency of 11.96\,GHz, is ($4.7 \pm 0.7) \times 10^7$\,K, implying a non-thermal emission mechanism.
The weak extension was fit by an unresolved component with a flux density of $ 84.2 \pm 27.5$\,$\mu$Jy ($\sim 3.1\sigma$).
We note that Gaussian components, as reported here, provide a convenient measure of source
structure even if they do not necessarily represent discrete physical structures.

Figure~2 shows the VLBA image of the \quasar\ at 4.67\,GHz ({\it Left}), and 7.67\,GHz ({\it Right}). Their restoring beams are $3.64 \times 1.62$~mas ($19.3 \times 8.6$\,pc) and $2.39 \times 0.98$\,mas ($12.7 \times 5.2$\,pc), respectively. The plus sign in each image denotes the location of the dominant source detected at 1.53\,GHz with the VLBA. The 3$\sigma$ point source upper limits are 51.6 and 61.2\,$\mu$Jy at 4.67 and 7.67\,GHz, respectively. The corresponding 3$\sigma$ upper limits to the intrinsic brightness temperatures are $3.8 \times 10^6$\,K, and $4.2 \times 10^6$\,K at the rest frame frequencies of 36.52 and 59.98\,GHz, respectively.

The limit on the spectral index between 1.53\,GHz and 4.67\,GHz for the dominant radio source seen in Figure~1 and using the 3$\sigma$ \added{point source} limit at 4.67\,GHz is $\alpha^{1.53}_{4.67} < -$1.55 (adopting $S\sim \nu^{\alpha}$). \added{Because the dominant radio source at 1.53\,GHz is resolved, this spectral index limit has been derived using its peak flux density value, which is $289.5\mu$\,Jy~beam$^{-1}$, on the assumption that this represents the maximum flux density of a point source at this frequency.} This derived value is steeper than that reported by \citet{ban21} between 1.52 and 2.87\,GHz with the VLA at a few arcsec resolution, which is $\alpha^{1.52}_{2.87} = -1.31$.

\subsection{The ``Companion'' Radio Source}

Figure~3 is the VLBA 1.53\,GHz image of the ``companion'' radio continuum source seen in the VLA images at a distance of 23\farcs1 from the quasar \quasar\ at 1.52 and 2.87\,GHz \citep{ban21}. The plus sign denotes the location of this radio source as seen in the VLA 2.87\,GHz image \citep{ban21}. The restoring beam size of the image shown in Figure~3 is $14.2 \times 5.9$~mas (P.~A.=$-18^{\circ}$), and the rms noise is 28.2~$\mu$Jy~beam$^{-1}$. No continuum emission is detected with the VLBA at the VLA position of this ``companion'' source with a 3$\sigma$ point source upper limit of 84.6\,$\mu$Jy, corresponding to an intrinsic brightness temperature upper limit of $4.1 \times 10^6$\,K at the rest frame frequency of 11.96\,GHz if located at $z=6.82$. At 1.53 GHz, the VLBA short spacing limit filters
out all spatial structures larger than about 0\farcs17, and the non-detection of this source at 1.53\,GHz is in agreement with its measured deconvolved size with the VLA at 2.87\,GHz, which is $1\farcs3 \times 0\farcs8$ \citep{ban21}.

\begin{figure}
\epsscale{1.02}
\plotone{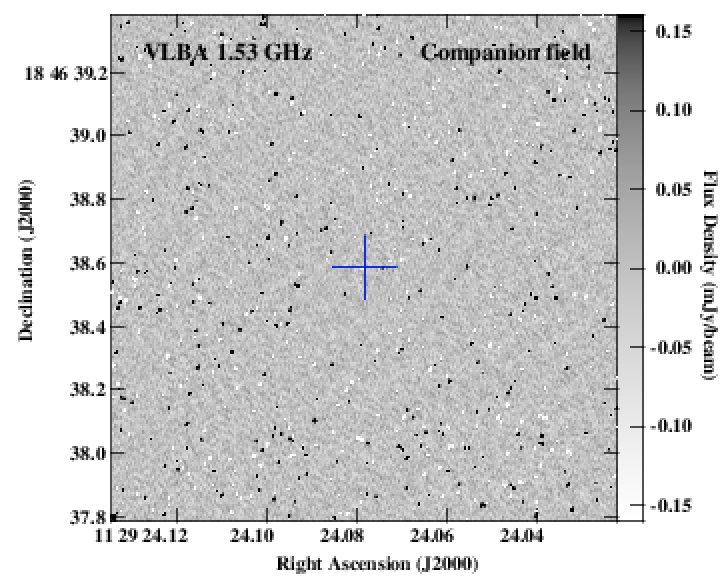}
\caption
{VLBA continuum image of the radio ``companion'' source of the $z=6.82$ quasar \quasar\ at 1.53\,GHz and $14.2 \times 5.9$~mas resolution (P.~A.=$-18^{\circ}$) located at an angular distance of 23\farcs1 from the quasar. The contour levels are at $-3$ and 3 times the rms noise level, which is 28.2~$\mu$Jy~beam$^{-1}$. The gray-scale range is indicated by the step wedge at the right side of the image. \added{The size of the image is $1\farcs6 \times 1\farcs6$, comparable to the deconvolved size limit of the source measured with the VLA at 1.52\,GHz.} The plus sign (arbitrary scale) denotes the VLA radio position at 2.87 GHz:
R.A.\,(J2000)=\,11$^{\rm h}$29$^{\rm m}$24.0782$^{\rm s}$, Decl.\,(J2000)=\,$+$18$^\circ$46$^\prime$38\farcs585.
\label{fig:vlba3}}
\end{figure}

Similar to the 1.53\,GHz results, both the 4.67\,GHz and 7.67\,GHz images do not show any radio emission from this ``companion'' source. The 3$\sigma$ point source upper limits are 53.4 and 62.4\,$\mu$Jy at 4.67 and 7.67\,GHz, respectively, and the corresponding upper limits to the intrinsic brightness temperatures are $4.0 \times 10^6$\,K, and $4.4 \times 10^6$\,K at the rest frame frequencies of 36.52 and 59.98\,GHz, respectively, if located at $z=6.82$.

\section{Discussion} \label{disc}

We have detected the $z=6.82$ quasar \quasar\ at 1.53~GHz with the VLBA at mas resolution (Figure\,1). The observations show that the radio emission from this source is dominated by a resolved compact source with a deconvolved size of $9.9 \times 3.5$\,mas ($52.5 \times 18.6$\,pc). A weak ($\sim 3.1\sigma$) extension to this dominant source is also seen in the VLBA image (Figure~1), but future observations are needed to unambiguously confirm its nature. The total radio flux density measured with the VLBA at 1.53\,GHz is $489.2 \pm 67.7$\,$\mu$Jy. This is consistent with the VLA measured value at 1.52\,GHz and $3\farcs55 \times 3\farcs24$ resolution, which is $510 \pm 15$\,$\mu$Jy \citep{ban21}, and suggests that the radio emission from the quasar is confined to the structures seen in the VLBA observations.

We find no indication of multiple radio components in the field of this source on scales of 20\,mas to a few arcseconds. A similar conclusion was reached for several other high redshift radio-loud quasars (e.g., \citealt{FR03,MOM04}).
These results imply that these quasars are not strongly gravitationally lensed.

The ``companion'' source identified in the VLA observations of \citet{ban21} and located at an angular distance of 23\farcs1 from the quasar was not detected with the VLBA at any of the three observed frequencies. While its nature and/or exact association with the quasar remains unclear, its non-detection in the VLBA observations at 1.53\,GHz suggests a size larger than 0\farcs17, which is expected for sources at lower redshift.

At the higher observing frequencies, namely 4.67 and 7.67\,GHz, the quasar itself was also not detected with the VLBA, perhaps contrary to expectations per the measured spectral index between
1.52 and 2.87\,GHz with the VLA, which is $\alpha^{1.52}_{2.87} = -1.31 \pm 0.08$ \citep{ban21}.

The spectral index limit we measure between 1.53 and 4.67\,GHz, which is $\alpha^{1.53}_{4.67} < -$1.55 (see \ref{raqso}), suggests a spectral steepening at the higher frequencies, while the slight flattening of the spectral index between the lower frequencies (1.52 and 2.87\,GHz) may be an indication of a spectral turnover taking place at much lower (few 100\, MHz) frequencies in the observed frame, which correspond to a few GHz in the rest-frame of the quasar.

The compact nature of this quasar located in the Epoch of Reionization, and its spectral trend speculated in the above at lower (few 100\, MHz) frequencies, would make it an important candidate for future sensitive H {\footnotesize I} 21\,cm absorption observations to detect the neutral IGM near $z \sim 7$ \citep{cgo02,FL02,GM17}. Knowledge of the source structure, as presented in these high angular resolution observations, is critical to identify candidates to search for H {\footnotesize I} and for subsequent interpretation of the results at these redshifts.

In the following we compare \quasar, which is the highest redshift radio-loud quasar known-to-date, with other radio loud quasars within a $\Delta z \sim \pm 1$, or $ z > 5.8$. In total, and including the quasar \quasar, there are currently nine radio-loud quasars at $z>5.8$. Following \citet{ban21}, we define $R_{2500}=f_{\nu,5\,\mathrm{GHz}} / f_{\nu,2500\,\mathrm{A}}$, with quasars that have $R_{2500} >10$ being radio loud\footnote{The radio-loudness of a quasar is defined as the ratio of the rest-frame 5\,GHz (radio) and 4400\,\AA\ (optical) flux densities; $R_{4400}$ \citep{kel89}, or the 2500\,\AA\ (ultra-violet) emission instead of the optical one; $R_{2500}$ \citep{jia07}.}. Figure~4 shows the rest-frame 5\,GHz radio vs.~the 2500\,\AA~luminosities of all the $z>5.8$ quasars that can be robustly classified as radio-quiet or radio-loud (see Table\,6 in \citealt{ban21}, as well as \citealt{ban15}, \deleted{and} \citealt{liu21}, \added{and \citealt{ighi21}}). 
The dashed lines represent the radio-to-optical ratios ($R_{2500}$) at three different values: 10, 100, and 1000. The quasar \quasar\ is denoted with the star sign and has a value of $R_{2500} = 91 \pm 9$ \citep{ban21}. Of the other eight radio-loud quasars in this $z>5.8$ sample, currently six have reported mas resolution radio imaging results through VLBI (marked by the filled circles in Figure\,4): J0836+0054 at $z=5.81$ \citep{FR05}, P352$-$15 at $z=5.84$  \citep{MOM18}, J2228+0110 at $z=5.95$ \citep{CAO14}, PSO J030947.49+271757.31 at $z=6.10$ (hereafter PSO J0309+27; \citealt{spi20}), J1427+3312 at $z=6.12$ \citep{FR08,MOM08}, and J1429+5447 at $z=6.18$ \citep{FR11}. VLBI imaging shows three of these sources are resolved into two or more distinct radio components with linear projected separations or extents of 174\,pc (J1427+3312), 500\,pc (PSO J0309+27) and 1.62\,kpc (P352$-$15). These sources have been interpreted as compact or medium-size symmetric objects (J1427+3312, P352$-$15), or one sided core-jets (P352$-$15\footnote{The exact nature of P352$-$15 is currently not clear. Based on single frequency VLBI observations, it has been interpreted as both classes of objects \citep{MOM18}. Future multi-frequency VLBI analysis would address this ambiguity (Momjian et al., in prep).}, PSO J0309+27). The other three quasars imaged with VLBI, J0836+0054, J2228+0110, and J1429+5447, are dominated by single compact sources on VLBI scales. Of these, J0836+0054 and J1429+5447 are also known to have steep spectra ($\alpha^{1.4}_{5.0} \leq -0.8$; \citealt{FR05,FR11}).

\begin{figure}
\epsscale{1.1}
\plotone{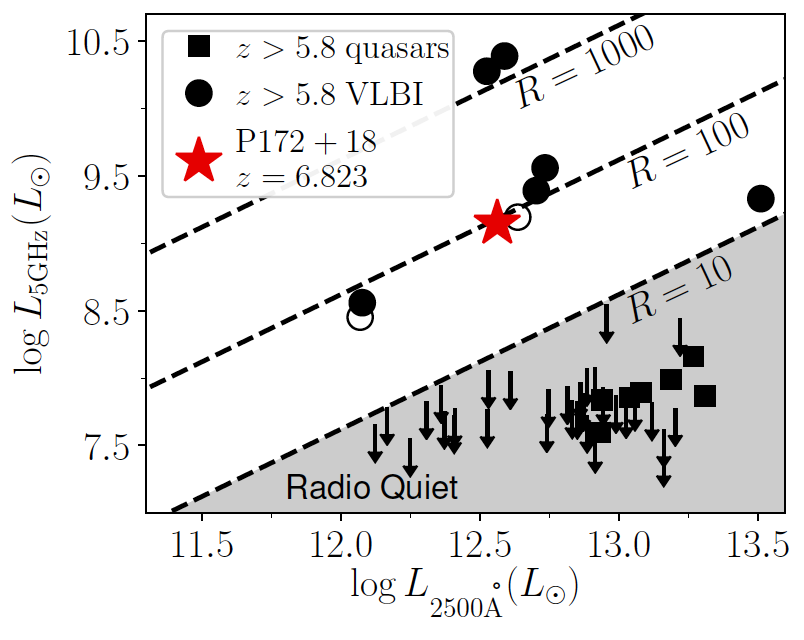}
\caption
{The 5\,GHz radio vs.\ the 2500\,\AA\ ultra-violet luminosities for the most distant quasars with redshifts $z>5.8$. The dashed lines represent the radio-loudness ratios ($R_{2500}$) at three different values: 10, 100, and 1000. Quasars with values of $R_{2500} >10$ are radio loud. The quasar \quasar\ at $z=6.82$ is denoted with the star sign and has a value of R $\sim 90$ \citep{ban21}. The open circles denote quasars that currently do not have published VLBI results: J2242+0334 at $z=5.88$ \citep{liu21} and VIK~J2318$-$3113 at $z=6.44$ \citep{ighi21}. The downward pointing arrows denote $z>5.8$ quasars with no radio detection. Their 5\,GHz luminosities are derived using $3\sigma$ limits from measurements at radio wavelengths \citep{ban21}.}
\end{figure}

To further investigate the nature of the $z=6.82$ quasar \quasar, we calculate the magnetic field strength and pressure in its dominant radio source adopting the standard minimum energy assumption for the fields and relativistic particle distribution (see the equations in \citealt{m80}). We assume equal energy in relativistic electrons and protons, a filling factor of unity, and a source size given by the VLBA observations. The derived minimum energy magnetic field is then 0.037\,G, and the energy density is $1.3\times 10^{-4}$ erg cm$^{-3}$. This field strength is comparable to those derived from VLBI observations of pc-scale jets in lower redshift radio AGN \citep{OG09}.

If the fields are close to the value derived assuming minimum energy, then the relativistic electron radiative lifetimes are short, $\le 0.7$\,years, using an upper limit to the synchrotron break frequency of 4.7\,GHz \citep{m80}. We have included both synchrotron losses and inverse Compton scattering off the CMB, although the former dominates, even at this redshift, given the high field strength. The source size of $\sim 50$\,pc implies a light crossing time of about 170\,years. Even if the jet is propagating at close to the speed of light, particle reacceleration in jet shocks appears to be required to maintain the electron distribution.

This source, with its steep spectral index and compactness, fits into the class of compact radio sources known as Gigahertz Peaked Spectrum (GPS) sources that have projected linear sizes less than 500\,pc \citep{ODea98,OS20}. Such sources, which host radio AGN, along with their more extended/evolved counterparts known as Compact Steep Spectrum radio sources (CSS; projected linear sizes of 0.5--20\,kpc), have been extensively studied at high angular resolution at lower redshifts (\citealt{OS20}, and references therein). One hypotheses to explain such sources is a very young radio jet. Assuming a median expansion speed for GPS/CSS sources of $\sim 0.1c$, the resulting kinematic age is 1700 years for the quasar \quasar, which is well within the range derived for lower redshift GPS/CSS sources \citep{OS20}. The extreme pressures in the radio emitting regions will certainly have a dramatic feedback effect on the ISM in the inner regions of the host galaxy, and continue to do so as the source expands \citep{ACF12}. Based on the above arguments, our conclusion of the $z=6.82$ quasar being similar to lower redshift GPS sources is also consistent with that reached for other $z\sim6$ single-source-dominated radio-loud quasars (e.g., \citealt{FR05, FR11}), implying these are very young radio sources in the early Universe.

\acknowledgments
The National Radio Astronomy Observatory is a facility of the National Science Foundation operated under cooperative agreement by Associated Universities, Inc. This work made use of the DiFX software correlator developed at Swinburne University of Technology as part of the Australian Major National Research Facilities program.

\facilities{VLBA}
\software{AIPS: \citealt{G2003}}

\listofchanges
\end{document}